# Infrared-Transparent Visible-Opaque Fabrics for Wearable Personal Thermal Management

Jonathan K. Tong[1,‡], Xiaopeng Huang[1,‡], Svetlana V. Boriskina[1,*], James Loomis[1], Yanfei Xu[1], and Gang Chen[1,*]

[1]Department of Mechanical Engineering, Massachusetts Institute of Technology, Cambridge, MA 02139

## Abstract

Personal cooling technologies locally control the temperature of an individual rather than a large space, thus providing personal thermal comfort while supplementing cooling loads in thermally regulated environments. This can lead to significant energy and cost savings. In this study, a new approach to personal cooling was developed using an infrared-transparent visible-opaque fabric (ITVOF), which provides passive cooling via the transmission of thermal radiation emitted by the human body directly to the environment. Here, we present a conceptual framework to thermally and optically design an ITVOF. Using a heat transfer model, the fabric was found to require a minimum infrared (IR) transmittance of 0.644 and a maximum IR reflectance of 0.2 to ensure thermal comfort at ambient temperatures as high as 26.1$^\circ$C (79$^\circ$F). To meet these requirements, an ITVOF design was developed using synthetic polymer fibers with an intrinsically low IR absorptance. These fibers were then structured to minimize IR reflection via weak Rayleigh scattering while maintaining visible opaqueness via strong Mie scattering. For a fabric composed of parallel-aligned polyethylene fibers, numerical finite element simulations predict 1 μm diameter fibers bundled into 30 μm yarns can achieve a total hemispherical IR transmittance of 0.972, which is nearly perfectly transparent to mid- and far-IR radiation. The visible wavelength properties of the ITVOF are comparable to conventional textiles ensuring opaqueness to the human eye. By providing personal cooling in a form amenable to everyday use, ITVOF-based clothing offers a simple, low-cost solution to reduce energy consumption in HVAC systems.

## TOC Graphic

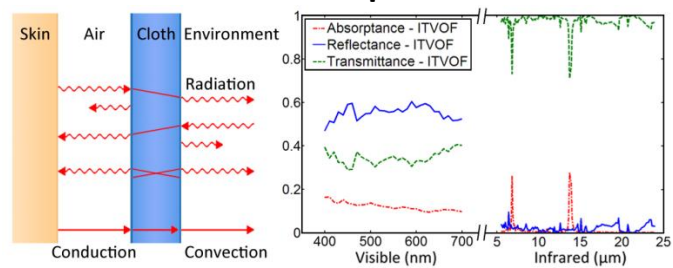

*Correspondence and requests for materials should be addressed to either S.V.B. or G.C. (emails: sborisk@mit.edu, gchen2@mit.edu)

In recent years, personal cooling technologies have been developed to provide local environmental control to ensure the user remains thermally comfortable when in extreme environmental conditions such as those faced by athletes, the military, or EMS personnel.[1] However, there remains a distinct lack of such technologies for everyday use by the average end user who spends the majority of the time in a sedentary state. This is especially important for indoor environments where incorporation of such technologies can offset energy consumed by HVAC systems for cooling while maintaining sufficient levels of thermal comfort. For instance, recent studies have shown that in the United States alone, residential and commercial buildings consume nearly 41% of total energy use each year with 37% of that energy devoted solely to heating and cooling.[2,3] To reduce energy usage, buildings have incorporated more renewable energy sources such as solar power, implemented advanced HVAC systems, utilized higher performing thermal insulation, and phase change materials for thermal storage all of which requires significant financial investment.[4–6] Instead, personal thermal comfort technologies offer a potentially low cost solution towards mitigating energy use by HVAC systems. Although these technologies can be used in a variety of indoor and outdoor environments, the focus of this work is to provide personal cooling in temperature regulated indoor environments.

At present, several technologies are commercially available which provide varying degrees of personal cooling. However, these technologies are typically tailored as high performance products, such as sportswear, body armor, and personal protection equipment, thus limiting functionality for everyday use. Arguably the most prevalent personal comfort technology used in industry today is moisture wicking where sensible perspiration is drawn away from the skin to the outer surface of the fabric and evaporated to ambient air thus cooling the wearer passively.[7–9] The drawback of this technology is that it is activated only when the wearer is sufficiently perspiring so that moisture accumulates on skin; thus, moisture wicking is not suitable to provide cooling for sedentary individuals. Other technologies utilize phase change materials in the form of cold packs which can effectively draw heat from the human body due to the high latent heat of melting associated with water and other refrigerants.[10–13] However this technology tends to be bulky in size and requires frequent replacement of the cold packs over time rendering this technology inconvenient and expensive to the end user. And finally, several technologies provide active cooling through use portable air conditioning units or liquid cooling.[14–18] These systems not only consume power, but also tend to be prohibitively expensive.

To overcome these limitations, we introduce the concept of an infrared-transparent visible-opaque fabric (ITVOF) which utilizes the human body's innate ability to thermally radiate heat as a cooling mechanism during the summer season when environmental temperatures are high. A heat transfer model was developed in order to determine the required IR optical properties of the ITVOF to ensure thermal comfort is maintained for environmental temperatures exceeding the neutral band. From this analysis, it was experimentally observed that existing textiles fail to meet these requirements due to a combination of intrinsic material absorption and structural backscattering in the IR wavelength range. In lieu of these loss mechanisms, a design for an ITVOF was developed using a combination of optimal material composition and structural photonic engineering. Specifically, synthetic polymers which support few vibrational modes were identified as candidate materials to reduce intrinsic material absorption in the IR wavelength range. To reduce backscattering losses, individual fibers were designed to be comparable in size to visible wavelengths in order to minimize reflection in the IR by virtue of weak Rayleigh scattering while remaining optically opaque in the visible wavelength range due to strong Mie scattering. By additionally reducing the size of the yarn, which is defined as a collection of fibers, less material is used thus decreasing volumetric absorption in the IR wavelength range even further. The ITVOF design is numerically demonstrated to exhibit a high transmittance and a low reflectance in the IR wavelength range while remaining optically opaque in the visible wavelength range. Compared to conventional technologies, an ITVOF can be manufactured into simple form factors while providing a fully passive means to cool the human body regardless of the physical activity level of the user.

## Heat Transfer Analysis

In order to quantify the potential cooling power using thermal radiation, the maximum radiative heat transfer achievable between the human body and the surrounding environment can be computed using the Stefan-Boltzmann law. Past studies have shown that human skin behaves like a blackbody with an emittance near unity in the IR wavelength range.[19,20] Even if the skin is wet due to perspiration, the emittance is still 0.96 corresponding to water, which suggests human skin is an effective IR emitter for all levels of physical activity.[21] If it is assumed the surface area of an average adult human body is A = 1.8 m$^2$, the temperature of human skin is $T_0$ = 33.9°C (93°F), and the ambient temperature is $T_3$ = 23.9°C (75°F), which corresponds to the upper limit of a typical neutral temperature band for human thermal comfort in buildings, the radiative heat transfer coefficient between the skin and the environment is $h_r$ = 6.25 W/m$^2$K.[22,23] Under these conditions, the cooling power predicted by the Stefan-Boltzmann law is 112 W.[24] Radiative heat loss from the human body is thus comparable to natural convection and the cooling power actually exceeds the total heat generation rate of $q_{gen}$ = 105 W assuming a base metabolic rate at rest of 58.2 W/m$^2$.[25] From this estimation, it can be observed that thermal radiation clearly has the potential to provide significant cooling power.

To fully harness thermal radiation for cooling, clothing fabrics should be transparent to mid- and far-infrared radiation which is the spectral range where the human body primarily emits.[19,24] Although a total hemispherical transmittance of unity would be ideal, it would be useful to determine the transmittance required for the ITVOF to provide the necessary cooling power for an individual to feel comfortable at different indoor temperatures. This criterion is determined by assuming the cooling power should equal the total heat generation rate of $q_{gen}$ = 105 W at a skin temperature of $T_0$ = 33.9°C (93°F) and a typical room temperature of $T_3$ = 23.9°C (75°F). Under these conditions, the effective heat transfer coefficient is equal to $h_{ref}$ = 5.8 W/m$^2$K which is less than the maximum achievable using thermal radiation. If the ambient temperature increases, the additional cooling power, $q_{cool}$, needed is equal to the difference between the total heat generation rate, $q_{gen}$, and the heat loss due to $h_{ref}$,

$$q_{cool} = q_{gen} - h_{ref}(T_0 - T_3) \qquad (1)$$

For this study, the goal is to provide cooling at an elevated ambient temperature of $T_3$ = 26.1°C (79°F), which past studies have shown can lead to nearly 40% energy savings in indoor environments for certain regions of the United States.[22] Using equation (1) at this temperature, the fabric must provide 23 W of additional cooling.

Based on this criterion, a more detailed one-dimensional steady-state heat transfer model is used to determine the total mid- to far-IR transmittance and reflectance required for the ITVOF. This model, as illustrated in Fig. 1, includes a continuous cloth placed at a distance, $t_a$, from the human body to model the effect of a thermally insulating air gap when loose fitting clothing is worn. The cloth is assumed to cover 100% of the human body. The model combines a control volume analysis and an analytical formulation of the temperature profile within the cloth in order to evaluate heat transfer between the human body, the cloth, and the ambient environment. Radiative heat transfer, heat conduction, and convection are all included in the analysis (see Supplementary Information for further details). The thermal conductivity of air is $k_a$ = 0.027 W/mK and the thermal conductivity of the cloth fabric is assumed to be $k_y$ = 0.05 W/mK.[24] Assuming a cloth porosity of 0.15, which is typical for common clothing, the effective thermal conductivity of the cloth layer is $k_c$ = 0.047 W/mK.[26] The thickness of the cloth is conservatively chosen to be $t_c$ = 0.5 mm and the effect of air circulation through the fabric is neglected. The IR optical properties of the cloth are assumed to be gray and diffuse with the sum of the absorptance, $\alpha_c$, reflectance, $\rho_c$, and transmittance, $\tau_c$, equal to 1.

From this model, the effect of the cloth's optical properties on the total cooling power was evaluated by calculating the maximum ambient temperature that can be sustained without compromising personal thermal comfort. Figures 2a and 2b show contour maps of the maximum ambient temperature as a function of the cloth's total reflectance and transmittance for different combinations of the air gap thickness, $t_a$, and the convective heat transfer coefficient, h, which represent a typical range of ambient environmental conditions where natural convection is dominant. In order to properly compare the impact of the cloth's optical properties on cooling for different environmental conditions, $t_a$ and h are coupled such that at an ambient temperature of 23.9°C (75°F) and assuming typical optical properties for clothing ($\rho_c \sim 0.3$ and $\tau_c \sim 0.03$), the total cooling power is always equal to the total heat generation rate thus ensuring a consistent baseline neutral temperature band is used.[27]

In both cases, a reflective cloth is more detrimental to cooling performance than an absorptive cloth since a high absorptance implies a high emittance, which would allow clothing to radiate thermal radiation to the environment albeit at a lower temperature. It can also be observed in Fig. 2a that in the limit of high absorption, the maximum ambient temperature is higher for the case where the air gap between the skin and cloth is less insulating ($t_a$ = 1.05 mm) despite the reduction in the convective heat transfer coefficient (h = 3 W/m²K). This suggests heat conduction and thermal radiation are comparable in this limit, thus for conventional clothing it is crucial to minimize the thermal resistance to heat conduction. However, as the mid- to far-IR transparency of the cloth increases, radiative heat transfer becomes more dominant compared to heat conduction. As a result, the impact of the insulating air gap on cooling is mitigated, thus a higher maximum ambient temperature can be sustained, when the convective heat transfer coefficient is larger (h = 5 W/m²K) even though the air gap is more insulating ($t_a$ = 2.36 mm) as shown in Fig. 2b. These results show that by designing clothing to be transparent to mid- and far-IR radiation, it is possible to provide persistent cooling using thermal radiation even for loose fitting clothing where the trapped air normally acts as a thermally insulating barrier, which impedes heat transfer in conventional personal cooling technologies.

To determine quantitatively the optical properties required for the ITVOF to provide 23 W of additional cooling at an ambient temperature of $T_3$ = 26.1°C (79°F), additional cooling power curves were computed as a function of the cloth's total reflectance and transmittance in Figs. 2c and 2d. In the limit of an ideal opaque fabric ($\alpha_c$ = 1), it can be seen for both cases that it is not possible to reach 23 W of additional cooling. This indicates that unless convective cooling is improved, which is challenging to achieve for everyday use as described earlier, it is impossible to maintain personal thermal comfort using opaque clothing; hence, the clothing must exhibit transparency to mid- and far-IR radiation.

For the case where $t_a$ = 1.05 mm and h = 3 W/m²K in Fig. 2c, if the cloth reflectance is larger than 0.2, it is also not possible to reach 23 W of additional cooling. This again shows that a higher cloth reflectance is more detrimental to the cooling performance of the cloth than absorption. Thus, when increasing the transmittance of the cloth, it is crucial that the reflectance is simultaneously reduced in order to maximize radiative cooling. Based on these results, the ITVOF must exhibit a maximum reflectance of 0.2 and a minimum transmittance of 0.644 in order to meet the personal thermal comfort criterion. For the case where $t_a$ = 2.36 mm and h = 5 W/m²K in Fig. 2d, the optical properties of the ITVOF become less stringent with a maximum reflectance of 0.3 and a minimum transmittance of 0.582. It should be emphasized that the reflectance and transmittance of the ITVOF are intrinsically coupled, thus a decrease in reflectance will lead to a corresponding decrease in the transmittance required to maintain thermal comfort as shown in Figs. 2c and 2d.

## Experimental Characterization of Common Clothing

In order to design the ITVOF, a baseline reference was first established by characterizing the optical properties of common clothing. Specifically, the optical properties of undyed cotton and polyester cloths, which comprise nearly 78% of all textile fiber production, were measured in both the visible and IR wavelength ranges.[28] Figures 3a and 3b show SEM images of the cotton and polyester cloths, respectively. The fabrics consist of fibers with a diameter of ~10 μm sewn into yarns that are 200 to 300 μm in size. Depending on the weave, the yarn can intertwine and overlap differently; however, the thickness of the cloth generally varies from one yarn to two overlapping yarns.

The visible wavelength optical properties of both cloth samples were measured using a UV/visible spectrometer. To account for the diffuse scattering of light from the samples, an integrating sphere was used to measure the hemispherical reflectance and transmittance of the cloth in the wavelength range of 400 nm to 800 nm. The results are shown in Fig. 3c. As expected, the undyed cloth samples exhibit no distinct optical features in the reflectance and transmittance spectra. Both samples show similar optical properties with a reflectance ranging from 0.4 to 0.5 and a transmittance ranging from 0.3 to 0.4. The high transmittance is primarily due to the intrinsic properties of cotton and polyester which are weakly absorbing in the visible wavelength range.[29,30]

Although these cloth samples exhibit a high transmittance, their apparent opaqueness is due to a combination of the contrast sensitivity of the human eye and the diffuse scattering of light. The human eye is a remarkably sensitive optical sensor that can respond to a large range of light intensities.[31,32] However, past studies have shown that the human eye can only perceive variations in light intensity when the change in intensity relative to the background is sufficiently large.[33–36] This implies that for clothing to appear opaque, the fraction of light reflected by the skin and observed by the human eye must be sufficiently smaller than the fraction of light reflected by the fabric into the same direction. For these cloth samples, light will reflect and transmit diffusively. In addition, skin is also a diffuse surface with a reflectance that is as high as 0.6 at longer wavelengths.[37] Since the observation of skin requires light to be reflected from the skin and transmitted through the cloth twice, more light will be scattered into directions beyond what is observable by the human eye compared to light that is only reflected by the cloth thus ensuring the opaque appearance of the cloth. It is for these reasons that common clothing appears opaque to the human eye despite an inherently high transmittance. From these results, the criteria for opaqueness of the ITVOF design will be assessed by comparing the hemispherical reflectance and transmittance to measured data shown in Fig 3c.

The IR transmittance spectra of the cloth samples, shown in Fig. 3d, were measured using a Fourier transform infrared (FTIR) spectrometer with a microscope objective accessory. Both the cotton and polyester samples exhibit a low transmittance of 1% across the entire IR wavelength range in agreement with previous studies.[27,38–40] Therefore, both samples are opaque in the IR and thus cannot provide the necessary cooling to the wearer at higher ambient temperatures according to the heat transfer model.

The reasons for the low transmittances are two-fold. First, cotton and polyester are highly absorbing in the IR wavelength range. Figure 4a shows the FTIR transmittance spectra of a single strand of cotton yarn and a polyester thin film. Several absorption peaks can be observed which originate from the many vibrational modes supported in the complex molecular structure of these materials. Since fabrics are typically several hundreds of microns thick, which is much larger than the penetration depth, incident IR radiation is completely absorbed at these wavelengths. Second, the fibers in clothing are comparable in size to IR wavelengths, as shown in Figs. 3a and 3b, which enable the fibers to support optical resonances that can strongly scatter incident light. In this Mie regime, it is well known that particles can exhibit large scattering cross sections due to these resonances.[41–46] For an array of many fibers, the collective scattering by the fibers can result in a high reflectance. Therefore, the creation of an ITVOF must minimize these two contributions in order to maximize transparency to mid- and far-IR radiation.

## Design and Simulation of an ITVOF

Based on the heat transfer modeling and the experimental results, the design strategy for an ITVOF is to use alternative synthetic polymers which are intrinsically less absorptive in the IR wavelength range and to structure the fibers to minimize the overall reflectance of the fabric in order to maximize radiative cooling. In general, synthetic polymers with simple chemical structures are ideal since fewer vibrational modes are supported thus resulting in less absorption. Additionally, these polymers must also be compatible with extrusion and drawing processes to ensure manufacturability for large scale production. Based on these criteria, polyethylene and polyacprolactam, more commonly known as nylon, were identified as potential candidate materials. It should be emphasized however that given the full gamut of synthetic polymers available, other synthetic polymers may also be suitable for an ITVOF.

Polyethylene is one of the simplest synthetic polymers available and the most widely used in industry today. The chemical structure of polyethylene consists of a repeating ethylene monomer with a total length that varies depending on the molecular weight. Because the chemical structure consists entirely of carbon-carbon and carbon-hydrogen bonds, few vibrational modes are supported. This is evidenced in Fig. 4b which shows measured FTIR transmittance spectra for an ultra-high molecular weight polyethylene (UHMWPE) thin-film (McMaster 85655K11). Absorption peaks can be observed at 6.8 µm corresponding to $CH_2$ bending modes, 7.3 µm and 7.6 µm to $CH_2$ wagging modes, and 13.7 µm and 13.9 µm to $CH_2$ rocking modes.[47] At longer wavelengths, additional rocking modes do exist, but are typically very weak. For textile applications, woven polyethylene fabrics are often used as geotextiles, tarpaulins, and tapes.[48] To assess the suitability of polyethylene for clothing applications, further studies are needed to evaluate mechanical comfort and durability.

Nylon (McMaster 8539K191) exhibits a similar structure to polyethylene with the key difference being the inclusion of an amide chemical group. As shown in Fig. 4b, this results in additional vibrational modes from 6 µm to 8 µm and 13 µm to 14 µm corresponding to the various vibrational modes from the amide group.[49] Although nylon is absorptive over a larger wavelength range compared to polyethylene, the advantage of nylon is that it is currently used in many textiles.

Compared to cotton and polyester, Fig. 4b shows that polyethylene and nylon exhibit fewer vibrational modes particularly in the mid-IR wavelength range near 10 µm where the human body thermally radiates the most energy. This indicates that polyethylene and nylon are intrinsically less absorptive and are therefore suitable for the creation of an ITVOF.

In order to further improve the IR transparency of an ITVOF constructed from these materials, structural photonic engineering can be introduced for both the fiber and yarn. Specifically, absorption by weaker vibrational modes can be minimized by reducing the material volume. This can be accomplished by simply decreasing the yarn diameter. To minimize backscattering of IR radiation, the fibers can also be reduced in size such that the diameter is small compared to IR wavelengths. In this manner, incident IR radiation will experience Rayleigh scattering where the scattering cross section of infinitely long cylinders in this regime decreases rapidly as a function of the diameter raised to the 4th power.[41] By reducing the scattering cross section, back scattering of IR radiation will significantly decrease resulting in an overall lower IR reflectance.

Conversely, for the visible wavelength range the ITVOF must instead have a low transmittance to ensure ITVOF-based clothing is opaque to the human eye. Since polyethylene and nylon are not strongly absorptive in the visible wavelength range, reflection must be maximized. This can be achieved by using fibers that are comparable in size to visible wavelengths so that incident light experiences Mie scattering. In exactly the same manner that conventional clothing is opaque to IR radiation, fibers in this regime can support optical resonances that significantly increase the scattering cross section of each fiber thus increasing the overall backscattering of incident light.[41] Since the fabric is composed of an array of these fibers, not only will the total reflectance increase, but light scattering with the cloth will become more diffuse. In conjunction with the contrast sensitivity of the human eye, this design approach can ensure the ITVOF is opaque to the human eye. Thus, the beauty of this structuring approach is that with an optimally chosen fiber diameter, two different regimes of light scattering are utilized in different spectral ranges in order to create a fabric which is simultaneously opaque in the visible wavelength range and transparent in the IR wavelength range.

In regards to the coloration of the ITVOF, polyethylene and nylon exhibit dispersionless optical properties in the visible wavelength range and with sufficient backscattering appear white in color. Despite the chemical inertness of polyethylene, it is possible to provide coloration through the introduction of pigments during fiber formation when polyethylene is in a molten state.[50] On the other hand, nylon fibers can be colored easily using conventional dyes.[51] Depending on the pigment or dye, additional vibrational modes may be introduced in the IR wavelength range reducing the overall transparency. However, for the sake of demonstrating the general concept of an ITVOF, this design aspect will be left for future studies.

To theoretically demonstrate the proposed strategy to create an ITVOF, numerical finite-element electromagnetic simulations were performed on a polyethylene fabric structure illustrated in Fig. 5. In these simulations, circular arrays of parallel fibers are arranged into collective bundles in order to represent the formation of yarn. The yarn is then positioned in a periodic staggered configuration oriented 30° relative to the horizontal plane to mimic the cross section of a woven fabric. For all simulations, it is assumed the fiber separation distance, $D_s$, is 1 μm and the yarn separation distance, $D_p$, is 5 μm, which is consistent with the cloth structures observed in Figs. 3a and 3b. Simulations in the IR wavelength range were conducted from 5.5 to 24 μm. Wavelengths shorter than 5.5 μm contribute only 2.7% to total blackbody thermal radiation and are thus considered negligible. Wavelengths longer than 24 μm contribute 17.2% to total blackbody thermal radiation; however, longer wavelengths are expected to yield an even higher transparency since polyethylene does not support vibrational modes beyond 24 μm. As a conservative estimate, the optical properties of the ITVOF design are spectrally integrated and normalized within only the 5.5 to 24 μm wavelength range, which will underestimate the transmittance and overestimate the reflectance and absorptance. Furthermore, the spectral integration is weighted by the Planck's distribution assuming a skin temperature of 33.9°C (93°F).

Floquet periodic boundary conditions are used on the right and left boundaries to simulate an infinitely wide structure. Perfectly matched layers are used on the top and bottom boundaries to simulate an infinite free space. Simulations were conducted for incident light polarized parallel and perpendicular to the fiber axis at normal incidence. The optical properties for unpolarized light were determined by taking an average of the results for both polarizations. The optical constants of bulk polyethylene were taken from the literature.[30] Although the manufacture of polymer fibers and the subsequent stress imposed when woven into fabrics can introduce anisotropy in the dielectric permittivity, past studies have experimentally shown that the optical properties of drawn UHMWPE exhibit minimal change when subjected to a high draw ratio and high stresses.[52,53] Therefore, anisotropic effects were neglected in this study.

To assess the impact of reducing the size of the fiber ($D_f$) and the yarn ($D_y$), the IR optical properties were computed by varying the yarn diameter ($D_y$ = 30 μm, 50 μm, and 100 μm) assuming a fixed fiber diameter of $D_f$ = 10 μm and by varying the fiber diameter ($D_f$ = 1 μm, 5 μm, and 10 μm) assuming a fixed yarn diameter of $D_y$ = 30 μm. The results are shown in Fig. 6 along with the total spectrally integrated IR transmittance ($\tau_c$) and reflectance ($\rho_c$) weighted by the Planck's distribution assuming a skin temperature of 33.9°C (93°F). Based on these results, a reduction in the yarn diameter does yield a higher spectral transmittance in the IR wavelength range as evidenced by the increase in the total hemispherical IR transmittance from 0.48 for $D_y$ = 100 μm to 0.76 for $D_y$ = 30 μm. Simultaneously, the total hemispherical IR reflectance also decreases from 0.35 to 0.19. This can be explained by a reduction in the total material volume, which is defined in this study for a single yarn on a per unit depth basis with units of $\mu m^2$ corresponding to the cross section of the yarn. As the yarn diameter decreases from $D_y$ = 100 μm to $D_y$ = 30 μm, the material volume decreases from 4870 $\mu m^2$ to 550 $\mu m^2$, which results in less absorption. Additionally, a reduction in the yarn diameter will also decrease the number of fibers that can scatter incident IR radiation, which leads to less reflection. These results suggest that by decreasing only the yarn diameter to $D_y$ = 30 μm, it is possible to create an ITVOF that already exceeds the minimum transmittance of 0.644 and maximum reflectance of 0.2 required to maintain thermal comfort at an ambient temperature of 26.1$^o$C (79$^o$F) in Fig. 2c.

However, the spectral optical properties in Fig. 6 indicate there is still room to further improve the transparency of the ITVOF as absorption and reflection are still substantial particularly at shorter wavelengths. In this wavelength range, the size of the fiber ($D_f$ = 10 μm) is comparable to the wavelength and is thus in the Mie regime where the fiber can support cavity resonances that can couple to and scatter incident IR radiation. Since the mode density of these resonances is higher at shorter wavelengths, the overall reflectance of the fabric structure will be higher as well. By reducing the size of the fiber, the number of supported cavity resonances will decrease resulting in a lower reflectance. The total material volume will also be reduced further from 550 $\mu m^2$ for $D_f$ = 10 μm to 136 $\mu m^2$ for $D_f$ = 1 μm thus decreasing the absorptance even further.

When the fiber diameter is reduced to 5 μm, the absorptance exhibits a marginal decrease. On the other hand, the reflectance actually increases compared to the case where $D_f$ = 10 μm. This indicates that the fiber is still sufficiently large enough to support cavity resonant modes. Although there are fewer modes supported, as shown by the variation in reflectance, these modes become leakier for smaller size fibers thus resulting in a larger scattering cross section and a higher reflectance. As a result, there is little enhancement to the overall IR transmittance. Once the fiber diameter decreases to 1 μm, the reflectance dramatically decreases, which suggests the fiber is sufficiently small such that incident mid- to far-IR radiation will primarily experience Rayleigh scattering. In this Rayleigh regime, the fibers are too small to support cavity mode resonances thus reducing the reflection of IR radiation. Furthermore, the reduction in fiber size further reduces the total material volume again decreasing the absorptance. As a result, the total mid- to far-IR transmittance further increases from 0.76 for $D_f$ = 10 μm to 0.972 for $D_f$ = 1 μm, making the structure even more transparent to thermal radiation emitted by the human body. Simultaneously, the total mid- to far-IR reflectance decreases substantially from 0.19 to 0.021. By reducing the fiber diameter, the resulting improvements to the optical properties enable this ITVOF design to clearly surpass the requirements needed to provide 23 W of additional cooling at an ambient temperature of 26.1$^o$C (79$^o$F) based on Figs. 2c and 2d. It should again be noted that the calculated optical properties of the ITVOF only considered wavelengths from 5.5 to 24 μm. Since polyethylene is transparent at longer wavelengths, these results likely underestimate the total hemispherical transmittance and overestimate the total hemispherical reflectance and absorptance for mid- and far-IR radiation.

To assess the visible opaqueness, additional simulations were performed for the polyethylene-based ITVOF design assuming a constant refractive index of n = 1.5 and an extinction coefficient of k = 5·10$^{-4}$ based on literature values for the visible wavelength range from 400 nm to 700 nm.[30] These simulations were performed for the optimal design where $D_f$ = 1 μm, $D_y$ = 30 μm, $D_s$ = 1 μm, and $D_p$ = 5 μm. The results are shown in Fig. 7 along with the experimentally measured optical properties of undyed cotton and polyester cloth for comparison. The polyethylene-based ITVOF design exhibits a total hemispherical reflectance higher than 0.5 and a hemispherical transmittance less than 0.4 across the entire visible wavelength range, which is comparable to the optical properties of the experimentally characterized cotton and polyester cloth samples. The oscillatory behavior of the total hemispherical absorptance is indicative of whispering gallery and Fabry-Perot resonances supported in each fiber (see Supplementary Information) which confirms that light interaction is indeed in the Mie regime. As a result, these optical resonances provide strong backscattering to help ensure the fabric is opaque.

It can also be observed in Fig. 7 that the reflectance and transmittance do not follow the same trend as the absorptance. This can be attributed to the optical coupling of neighboring fibers in the fabric which collectively introduce additional optical resonances in the system due to the periodic nature of the assumed fabric structure. For a more realistic fabric structure where fiber and yarn spacing are nonuniform, long range optical coupling will be minimized resulting in a fabric which more diffusively scatters light. Due to the similarity to the experimentally characterized cloth samples, these results suggest the ITVOF design is optically opaque to the human eye. Furthermore, Fig. 7 clearly shows the contrast between the visible and IR properties of the ITVOF design which indicates that by optimally sizing the fiber, two vastly different regimes of light scattering can be simultaneously used.

Based on these results, it may appear that the creation of an ITVOF requires substantial reduction in material volume since the highest IR transmittance of 0.972 was predicted for the smallest yarn diameter ($D_y$ = 30 µm) and fiber diameter ($D_f$ = 1 µm). Although decreasing both of these parameters will certainly improve the overall transmittance in the IR wavelength range, additional simulations (see Supplementary Information) for various fiber diameters ($D_f$ = 1 µm, 5 µm, and 10 µm) assuming a larger yarn diameter of $D_y$ = 50 µm show a similar trend in the enhancement of transmittance. For a fiber diameter of $D_f$ = 1 µm, the total hemispherical IR transmittance and reflectance was 0.969 and 0.019, respectively, which is similar to the case where $D_y$ = 30 µm despite a material volume that is three times larger at 445 µm$^2$. This result shows that reducing the fiber diameter is far more crucial to improving transmittance compared to reducing the yarn diameter. Therefore, it may be suitable to create an ITVOF that is comparable in size to conventional fabrics so long as the fiber diameter is sufficiently small.

In addition, a polyethylene-based ITVOF may not exhibit sufficient fabric handedness due to the nature of the material used. To ensure the fabric is comfortable to the wearer, it may be necessary for the fabric to be composed of a mixture of different material fibers which will affect the transmittance of the fabric. To assess the potential extent in which the transmittance will be reduced, simulations were also performed (see Supplementary Information) for different volumetric concentrations of polyethylene and polyester (PET) again assuming $D_f$ = 1 µm and $D_y$ = 30 µm. The optical constants for PET were also taken from the literature.[29] For the most absorbing case of 25%PE/75%PET, the total hemispherical mid- to far-IR transmittance and reflectance was 0.728 and 0.038, respectively, which indicates that a fabric blend can still achieve a high transmittance and a low reflectance to provide sufficient cooling using thermal radiation.

In summary, we demonstrated a design for an infrared-transparent visible-opaque fabric (ITVOF) in order to provide personal cooling via thermal radiation from the human body to the ambient environment. We developed an ITVOF design made of polyethylene, which is an intrinsically low absorbing material, and structured the fibers to be sufficiently small in order to maximize the IR transparency and the visible opaqueness. For a 1 µm diameter fiber and a 30 µm diameter yarn, the total mid- and far-IR transmittance and reflectance are predicted to be 0.972 and 0.021, respectively, which exceed the minimum transmittance of 0.644 and maximum reflectance of 0.2 required to provide sufficient cooling at an elevated ambient temperature of 26.1°C (79°F). Simultaneously, the total hemispherical reflectance and transmittance in the visible wavelength range are comparable to existing textiles which indicates that the design is optically opaque to the human eye.

To practically realize an ITVOF, further studies are needed to experimentally evaluate the impact of radiative heat transfer on personal cooling. Although challenging, the fabrication of an ITVOF could be achieved using conventional manufacturing processes including drawing, extrusion, or electrospinning. Thermal and mechanical evaluation can be conducted using standardized testing methods as shown in previous studies including the use of thermal manikins, wash and dry cycling, and subject testing.[11,15,54–56] Additionally, vapor transport through the cloth, which is another key component for thermal comfort, must also be considered in future ITVOF designs. Although the porosity of the proposed ITVOF design is based on typical clothing, it would nonetheless be useful to quantitatively assess vapor transport to optimally design ITVOF-based clothing.[57] The inclusion of coloration for aesthetic quality is another important aspect that must be considered without compromising the effectiveness of radiative cooling. Alternative synthetic polymers, such as polypropylene or polymeric blends of UHMWPE and PET, should also be investigated for their suitability in an ITVOF design. Ultimately, ITVOF-based clothing offers a simple, low-cost approach to provide cooling locally to the human body in a variety of indoor and outdoor environments without requiring additional energy consumption, compromising breathability, or requiring any lifestyle change. Therefore, ITVOF provides a simple solution to reduce the energy consumption of HVAC systems by enabling higher temperature set points during the summer.

## Methods

**UV/Visible Characterization.** A custom UV/visible wavelength spectrometer was used to measure the optical properties of the cloth samples in the visible wavelength range. This system consisted of a 500 W mercury xenon lamp source (Newport Oriel Instruments, 66902), a monochromator (Newport Oriel instruments, 74125), an integrating sphere (Newport Oriel Product Line, 70672) and a silicon photodiode (Newport Oriel instruments, 71675). Total hemispherical reflectance measurements were performed by placing the cloth samples onto a diffuse black reference (Avian Technologies LLC, FGS-02-02c) to avoid reflection from the underlying substrate. Total hemispherical transmittance measurements were performed by placing the cloth samples onto the input aperture of the integrating sphere. All measurements were calibrated using a diffuse white reference (Avian Technologies LLC, FWS-99-02c).

**Infrared Characterization.** A commercially available FTIR spectrometer (Thermo Fisher Scientific, Nicolet 6700) and an IR objective accessory (Thermo Fisher Scientific, Reflachromat 0045-402) was used to measure the optical properties of the cloth samples and the polymer films in the infrared wavelength range. The objective was placed 15 mm behind the samples, corresponding to the working distance of the objective, in order to capture infrared radiation transmitted through the samples. For the cloth samples, the total hemispherical transmittance will be underestimated since not all of the IR radiation that is diffusively transmitted through the cloth sample is captured. However, the objective used in this study was designed to capture IR radiation at a 35.5° acceptance angle. Since it is expected that IR radiation will transmit diffusively, the measured results are likely underestimated by a few percent, which is still in agreement with previous studies.

## Associated Content

Supporting Information provides further details on the heat transfer modeling, the optical constants of polyethylene (PE) and polyethylene terephthalate (PET), Mie theory calculations for a single isolated polyethylene fiber, numerical finite element simulations of a polyethylene-based ITVOF for a larger yarn diameter, and numerical finite element simulations for an ITVOF blend of polyethylene and polyester.

## Author Information

**Corresponding Authors**
*E-mail: sborisk@mit.edu, gchen2@mit.edu
**Author Contributions**
‡ These authors contributed equally to this work.
**Notes**
The authors declare no competing financial interest.

## Acknowledgements
The authors would like to thank Wei-Chun Hsu, Yi Huang, and Lee Weinstein for helpful discussions. FTIR transmittance measurements were carried out at the Massachusetts Institute of Technology Center for Materials Science and Engineering (CMSE). This work was not supported by any federal research grants.

## References
(1) Yazdi, M.; Sheikhzadeh, M. Personal Cooling Garments: A Review. *J. Text. Insititute* **2014**, *105*, 1231–1250.
(2) 2011 Buildings Eenrgy Data Book. U.S. Department of Energy. Energy Efficiency & Renewable Energy Department (2011).
(3) Pérez-Lombard, L.; Ortiz, J.; Pout, C. A Review on Buildings Energy Consumption Information. *Energy Build.* **2008**, *40*, 394–398.
(4) Sadineni, S. B.; Madala, S.; Boehm, R. F. Passive Building Energy Savings: A Review of Building Envelope Components. *Renew. Sustain. Energy Rev.* **2011**, *15*, 3617–3631.
(5) Wang, S.; Ma, Z. Supervisory and Optimal Control of Building HVAC Systems: A Review. *HVAC&R Res.* **2008**, *14*, 3–32.
(6) Memon, S. A. Phase Change Materials Integrated in Building Walls: A State of the Art Review. *Renew. Sustain. Energy Rev.* **2014**, *31*, 870–906.
(7) Hong, C. J.; Kim, J. B. A Study of Comfort Performance in Cotton and Polyester Blended Fabrics. I. Vertical Wicking Behavior. *Fibers Polym.* **2007**, *8*, 218–224.
(8) Kaplan, S.; Okur, A. Thermal Comfort Performance of Sports Garments with Objective and Subjective Measurements. *Indian J. Fibre Text. Res.* **2012**, *37*, 46–54.
(9) Das, B.; Das, A.; Kothari, V. K.; Fanguiero, R.; de Araújo, M. Effect of Fibre Diameter and Cross-Sectional Shape on Moisture Transmission through Fabrics. *Fibers Polym.* **2008**, *9*, 225–231.
(10) McCullough, E. A.; Eckels, S. Evaluation of Personal Cooling Systems for Soldiers. In *13th Int. Environmental Ergonomics Conf.*; Boston, MA, USA, 2009; pp. 200–204.
(11) Gao, C.; Kuklane, K.; Wang, F.; Holmér, I. Personal Cooling with Phase Change Materials to Improve Thermal Comfort from a Heat Wave Perspective. *Indoor Ai* **2012**, *22*, 523–530.
(12) Muir, I. H.; Bishop, P. A.; Ray, P. Effects of a Novel Ice-Cooling Technique on Work in Protective Clothing at 28C, 23C, and 18C WBGTs. *Am. Ind. Hyg. Assoc. J.* **1999**, *60*, 96–104.
(13) Rothmaier, M.; Weder, M.; Meyer-Heim, A.; Kesselring, J. Design and Performance Cooling Garments Based on Three-Layer Laminates. *Med. Biol. Eng. Comput.* **2008**, *46*, 825–832.
(14) Elbel, S.; Bowers, C. D.; Zhao, H.; Park, S.; Hrnjak, P. S. Development of Microclimate Cooling Systems for Increased Thermal Comfort of Individuals. In *International Refrigeration and Air Conditioning Conference*; 2012; p. 1183.
(15) Kayacan, O.; Kurbak, A. Effect of Garment Design on Liquid Cooling Garments. *Text. Res. J.* **2010**, *80*, 1442–1455.
(16) Yang, J.-H.; Kato, S.; Seok, H.-T. Measurement of Airflow around the Human Body with Wide-Cover Type Personal Air-Conditioning with PIV. *Indoor Built Environ.* **2009**, *18*, 301–312.
(17) Yang, Y.-F.; Stapleton, J.; Diagne, B. T.; Kenny, G. P.; Lan, C. Q. Man-Portable Personal Cooling Garment Based on Vacuum Desiccant Cooling. *Appl. Therm. Eng.* **2012**, *47*, 18–24.
(18) Nag, P. K.; Pradhan, C. K.; Nag, A.; Ashetekar, S. P.; Desai, H. Efficacy of a Water-Cooled Garment for Auxiliary Body Cooling in Heat. *Ergonomics* **1998**, *41*, 179–187.
(19) Steketee, J. Spectral Emissivity of Skin and Pericardium. *Phys. Med. Biol.* **1973**, *18*, 686–694.
(20) Sanchez-Marin, F. J.; Calixto-Carrera, S.; Villasenor-Mora, C. Novel Approach to Assess the Emissivity of the Human Skin. *J. Biomed. Opt.* **2009**, *14*, 024006.

(21) Incropera, F. P.; Dewitt, D. P.; Bergman, T. L.; Lavine, A. S. *Fundamentals of Heat and Mass Transfer*; John Wiley & Sons, Inc., 2007.
(22) Hoyt, T.; Lee, K. H.; Zhang, H.; Arens, E.; Webster, T. Energy Savings from Extended Air Temperature Setpoints and Reductions in Room Air Mixing. *Int. Conf. Environ. Ergon. 2009* **2005**.
(23) Federspiel, C. Predicting the Frequency and Cost of Hot and Cold Complaints in Buildings. *Cent. Built Environ.* **2000**.
(24) Mills, A. F. *Heat Transfer*; Prentice Hall, 1998.
(25) *ASHRAE Handbook-Fundamentals*; ASHRAE, 2005.
(26) Jakšić, D.; Jakšić, N. Porosity of the Flat Textiles. In *Woven Fabric Engineering*; Sciyo, 2010; pp. 255–272.
(27) Lee, T.-W. *Thermal and Flow Measurements*; CRC Press, 2008.
(28) Oerlikon Leybold Group. The Fiber Year 2006/07 - A World Survey on Textile and Nonwovens Industry. **2007**.
(29) Laskarakis, A.; Logothetidis, S. Study of the Electronic and Vibrational Properties of Poly(ethylene Terephthalate) and Poly(ethylene Naphthalate) Films. *J. Appl. Phys.* **2007**, *101*, 053503.
(30) Palik, E. D. *Handbook of Optical Constants of Solids*; Academic Press, 1997.
(31) Ferwada, J. Elements of Early Vision for Computer Graphics. *IEEE Comput. Graph. Appl.* **2001**, *21*, 21–23.
(32) Wandell, B. A. *Foundations of Vision*; Sinauer Associates, 1995.
(33) Stevens, S. S. On the Psychophysical Law. *Psychol. Rev.* **1957**, *64*, 153–181.
(34) Fechner, G. T. *Elemente Der Psychophysik*; Breitkopf und Hartel: Leipzig, 1860.
(35) Stevens, S. S. To Honor Fechner and the Repeal of His Law. *Science* **1961**, *133*, 80–86.
(36) Steinhardt, J. Intensity Discrimination in the Human Eye: I. The Relation of DetlaI/I to Intensity. *J. Gen. Physiol.* **1936**, *20*, 185–209.
(37) Norvang, L. T.; Milner, T. E.; Nelson, J. S.; Berns, M. W.; Svaasand, L. O. Skin Pigmentation Characterized by Visible Reflectance Measurements. *Lasers Med. Sci.* **1997**, *12*, 99–112.
(38) Zhang, H.; Hu, T.; Zhang, J. Transmittance of Infrared Radiation Through Fabric in the Range 8-14 Mm. *Text. Res. J.* **2010**, *80*, 1516–1521.
(39) Carr, W. W.; Sarma, D. S.; Johnson, M. R.; Do, B. T.; Williamson, V. A.; Perkins, W. A. Infrared Absorption Studies of Fabrics. *Text. Res. J.* **1997**, *67*, 725–738.
(40) Xu, W.; Shyr, T.; Yao, M. Textiles' Properties in the Infrared Irradiation. *Text. Res. J.* **2007**, *77*, 513–519.
(41) Bohren, C. F.; Huffman, D. R. *Absorption and Scattering of Light by Small Particles*; WILEY-VCH Verlag GmbH & Co. KGaA, 2007.
(42) Bohren, C. F.; Huffman, D. R. Absorption and Scattering by an Arbitrary Particle. *Absorpt. Scatt. Light by Small Part.* **1998**, 57–81.
(43) Brönstrup, G.; Jahr, N.; Leiterer, C.; Csáki, A.; Fritzsche, W.; Christiansen, S. Optical Properties of Individual Silicon Nanowires for Photonic Devices. *ACS Nano* **2010**, *4*, 7113–7122.
(44) Cao, L.; White, J. S.; Park, J.-S.; Schuller, J. A.; Clemens, B. M.; Brongersma, M. L. Engineering Light Absorption in Semiconductor Nanowire Devices. *Nat. Nanotechnol.* **2009**, *8*, 643–647.
(45) Tong, J. K.; Hsu, W.-C.; Han, S.-E.; Burg, B. R.; Zheng, R.; Shen, S.; Chen, G. Direct and Quantitative Photothermal Absorption Spectroscopy of Individual Particulates. *Appl. Phys. Lett.* **2013**, *103*.
(46) Boriskina, S. V.; Sewell, P.; Benson, T. M.; Nosich, A. I. Accurate Simulation of 2D Optical Microcavities with Uniquely Solvable Boundary Integral Equations and Trigonometric-Galerkin Discretization. *J. Opt. Soc. Am. A* **2004**, *21*, 393–402.
(47) Krimm, S.; Liang, C. Y.; Sutherland, G. B. B. M. Infrared Spectra of High Polymers. II. Polyethylene. *J. Chem. Phys.* **1956**, *25*, 549.

(48) Crangle, A. Types of Polyolefin Fibres. In *Polyolefin Fibres: Industrial and Medical Applications*; Ugbolue, S., Ed.; Woodhead Publishing in Textiles, 2009; pp. 3–34.
(49) *Infrared and Raman Spectroscopy: Methods and Applications*; Schrader, B., Ed.; VCH, 2007.
(50) Charvat, R. A. *Coloring of Plastics: Fundamentals*; John Wiley & Sons, Ltd., 2005.
(51) *Colorants and Auxiliaries: Organic Chemistry and Application Properties*; Shore, J., Ed.; Society of Dyers and Colourists, 2002.
(52) Schael, G. w. Determination of Polyolefin Film Properties from Refractive Index Measurements. II. Birefringence. *J. Appl. Polym. Sci.* **1968**, *12*, 903–914.
(53) Wool, R. P.; Bretzlaff, R. S. Infrared and Raman Spectroscopy of Stressed Polyethylene. *J. Polym. Sci. Part B Polym. Phys.* **1986**, *24*, 1039–1066.
(54) ASTM D3995-14, Standard Performance Specification for Men's and Women's Knitted Career Apparel Fabrics: Dress and Vocational.
(55) ASTM Standard F1868, Standard Test Method for Thermal and Evaporative Resistance of Clothing Materials Using a Sweating Hot Plate.
(56) ISO 11092 Textiles - Physiological Effects - Measurement of Thermal and Water-Vapour Resistance under Steady-State Conditions (sweating Guarded-Hotplate Test).
(57) ASTM Standard E96/ E96M, 2013, Standard Test Methods for Water Vapor Transmission of Materials, 2013.

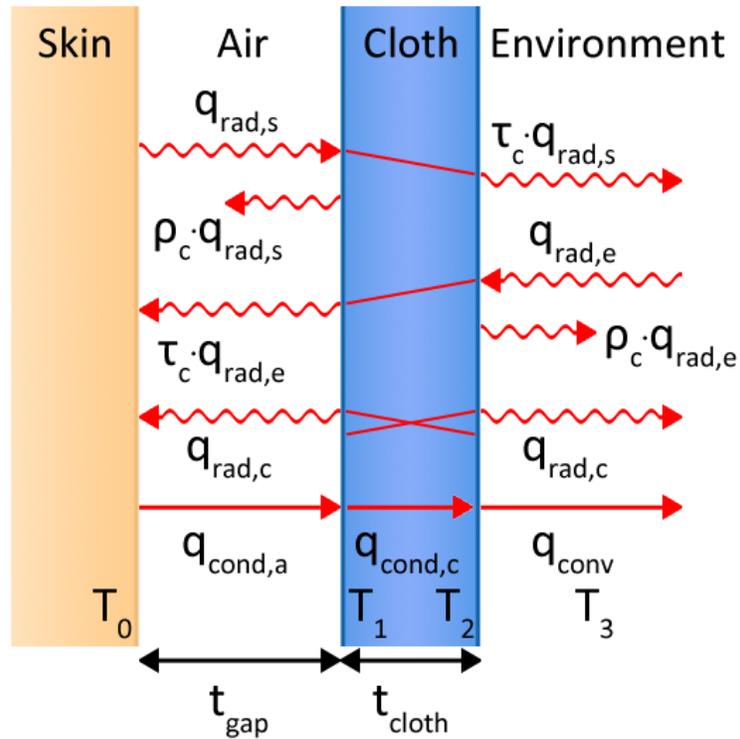

Figure 1: A heat transfer model was developed to analyze heat dissipation from a clothed human body to the ambient environment. Various heat transfer contributions that lead to dissipation of heat from the human body, such as radiation, heat conduction, and heat convection are included. To model loose fitting clothing, a finite air gap is assumed between the cloth and the skin.

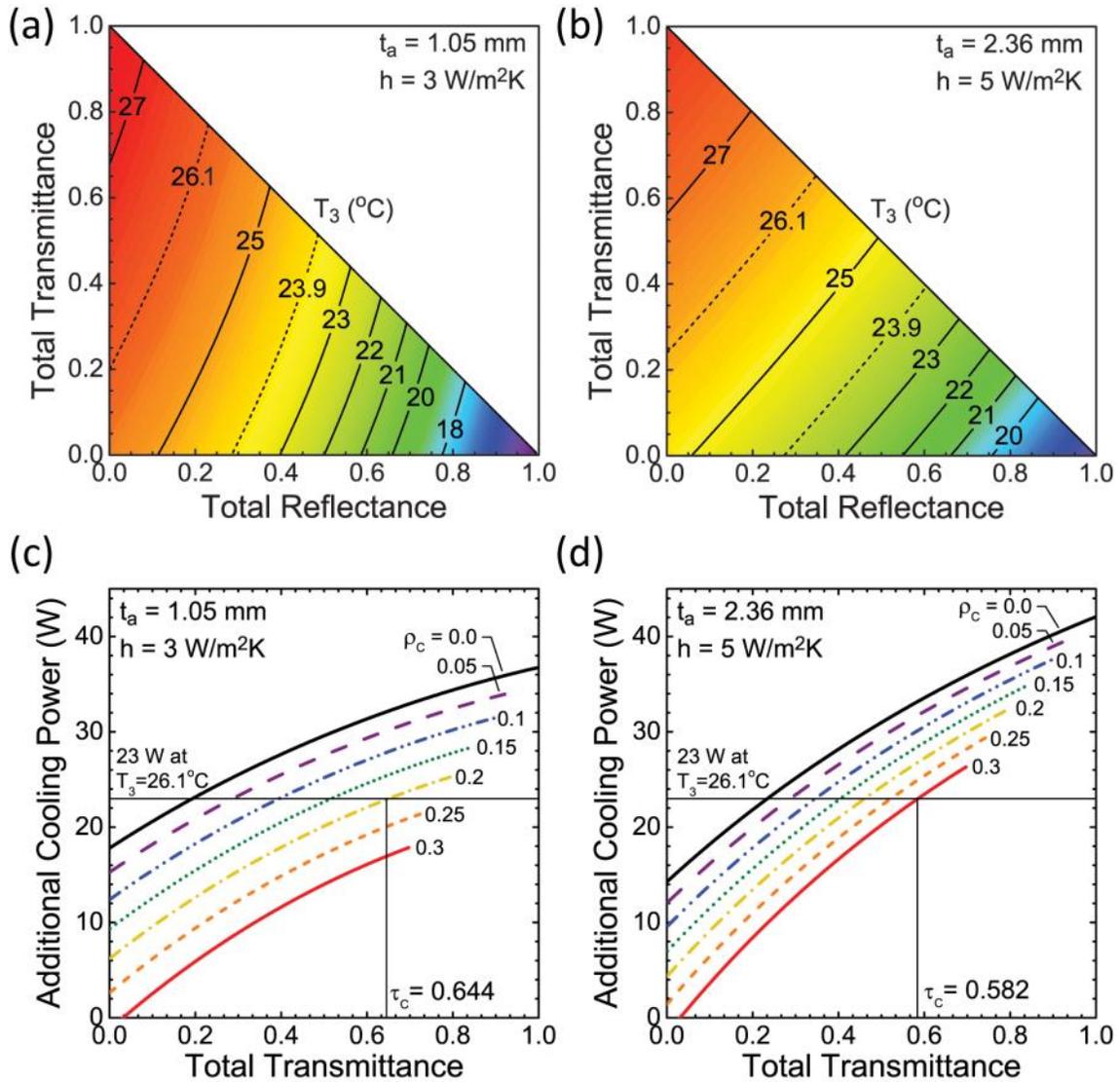

Figure 2: Evaluation of ITVOF mid- to far-IR optical requirements to maintain personal thermal comfort at elevated ambient temperatures. (a) A temperature map was computed showing the maximum ambient temperature attainable without compromising thermal comfort as a function of the total reflectance and transmittance of the cloth. It is assumed the air gap is $t_a$ = 1.05 mm and the convective heat transfer coefficient is h = 3 W/m²K. (b) A corresponding temperature map assuming $t_a$ = 2.36 mm and h = 5 W/m²K. The range of h is typical for cooling via natural convection. (c) An additional cooling power curve showing quantitatively the effect of radiative cooling as a function of the total cloth transmittance and reflectance assuming $t_a$ = 1.05 mm and h = 3 W/m²K. (d) An additional cooling power curve assuming $t_a$ = 2.36 mm and h = 5 W/m²K. As shown, by decreasing the reflectance and increasing the transmittance, it is possible to achieve the necessary 23 W of cooling at an ambient temperature of 26.1 °C using only thermal radiation.

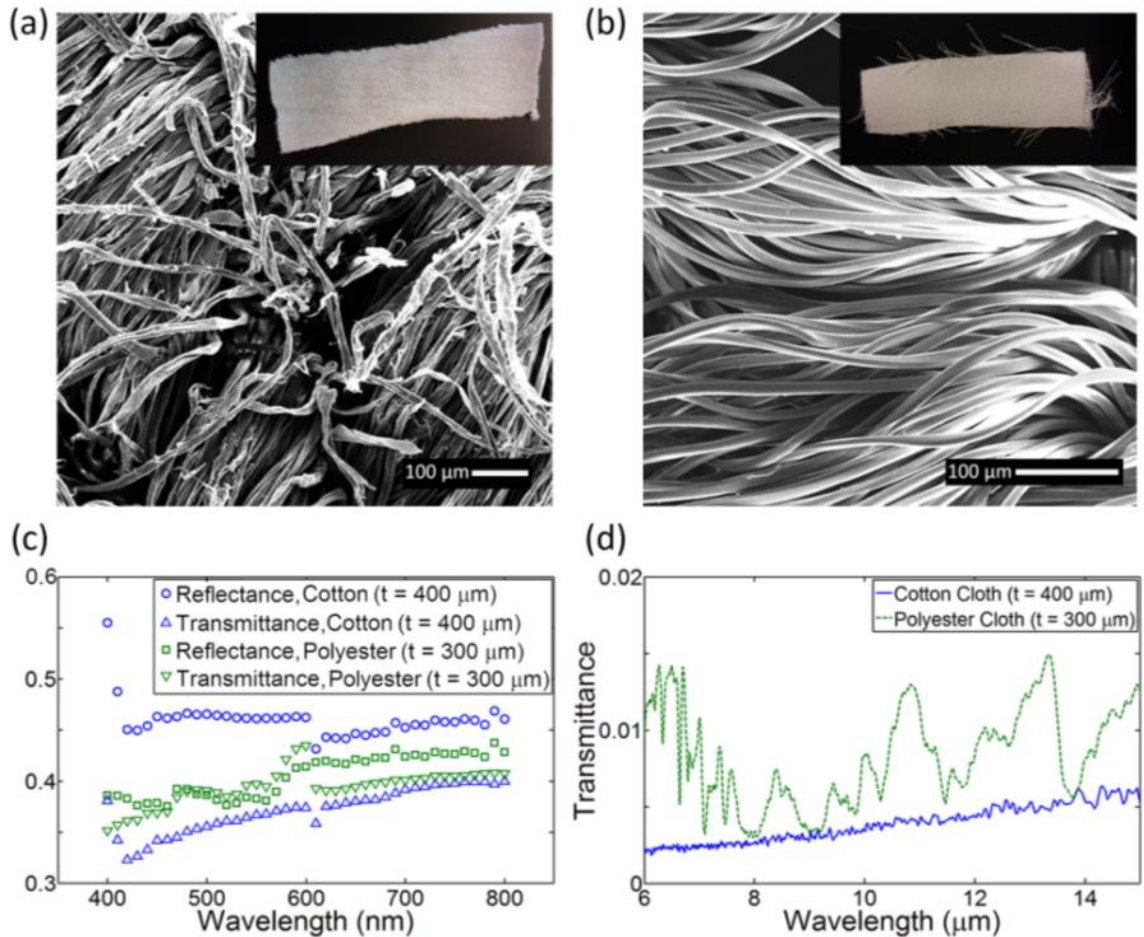

Figure 3: Optical properties of conventional clothing. SEM images of (a) undyed cotton cloth and (b) undyed polyester cloth which show the intrinsic fabric structure. The insets are optical images of the samples characterized. For both samples, the fiber diameter is on average 10 µm and the yarn diameter is greater than 200 µm. The scale bars both correspond to 100 µm. (c) Experimentally measured optical properties in the visible wavelength range. (d) Experimentally measured FTIR transmittance spectra of undyed cotton cloth (thickness, t = 400 µm) and undyed polyester cloth (t = 300 µm) showing the opaqueness of common fabrics in the IR.

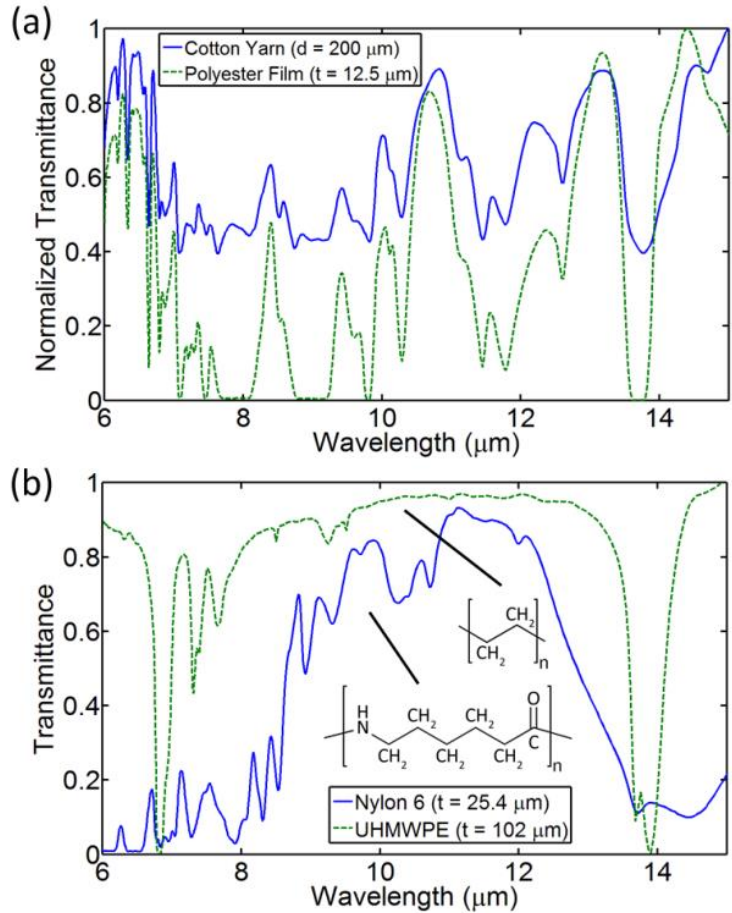

Figure 4: Intrinsic absorptive properties of various synthetic polymers. (a) The FTIR transmittance spectra for a single cotton yarn (diameter, d = 200 µm) and a polyester thin-film (thickness, t = 12.5 µm). The transmittance spectra is normalized to provide similar scaling due to the order of magnitude difference in sample size. (b) The FTIR transmittance spectra for two candidate materials for the ITVOF. These materials include thin-films of nylon 6 (t = 25.4 µm) and UHMWPE (t = 102 µm).

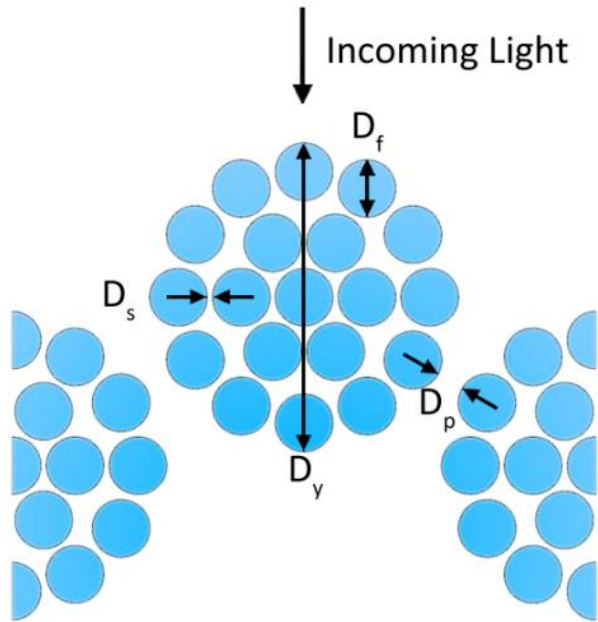

Figure 5: A schematic of the numerical simulation model used to predict the optical properties of the ITVOF design. The parameters include: $D_f$ – the fiber diameter, $D_y$ – the yarn diameter, $D_s$ – the fiber separation distance, and $D_p$ – the yarn separation distance. For all simulations, the yarns were staggered $30°$ relative to the horizontal plane. In addition, incident light was assumed to be at normal incidence and the optical properties for unpolarized light were calculated by average light polarized parallel and perpendicular to the fiber axis.

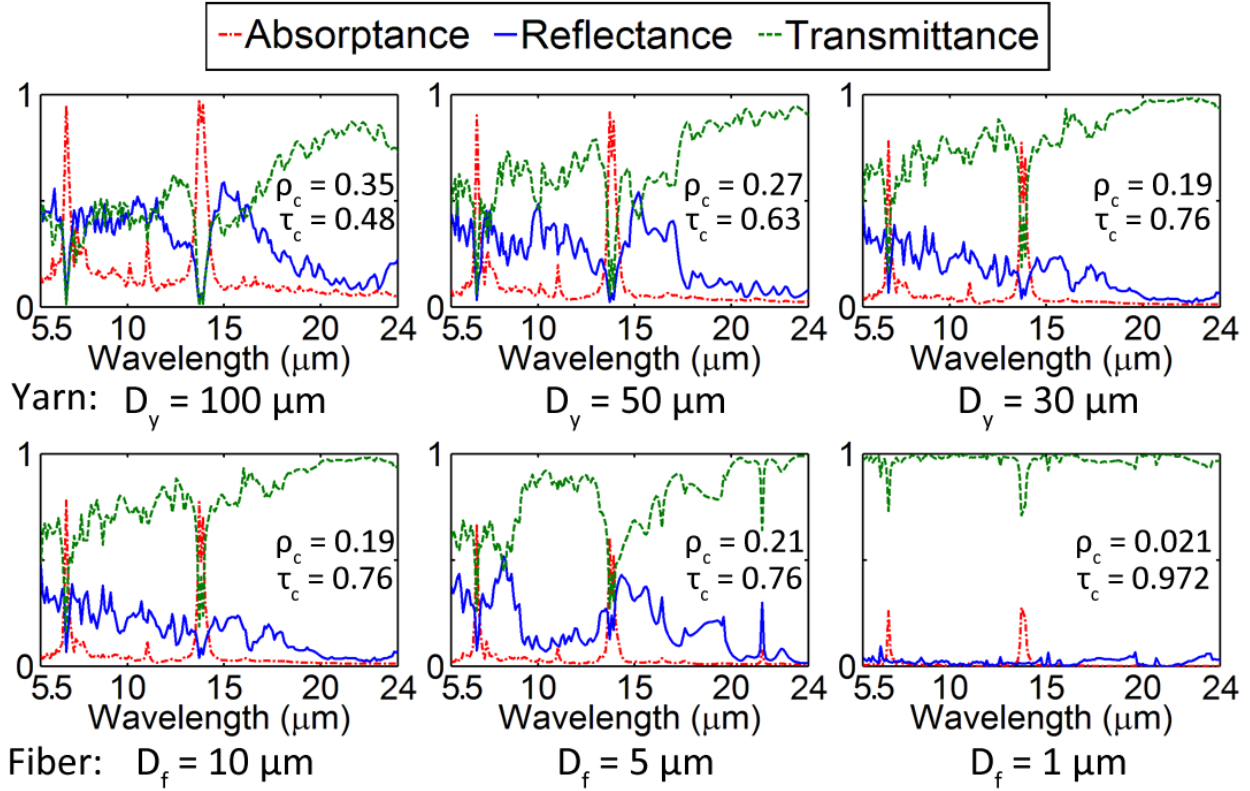

Figure 6: Numerical simulation results for the IR optical properties of a polyethylene-based ITVOF illustrating the effect of reducing the fiber and yarn size. Upper row: The yarn diameter is varied ($D_y$ = 30 µm, 50 µm, and 100 µm) assuming a fixed fiber diameter of $D_f$ = 10 µm. Lower row: The fiber diameter is varied ($D_f$ = 1 µm, 5 µm, and 10 µm) assuming a fixed yarn diameter of $D_y$ = 30 µm. For all simulations, the fiber separation distance is $D_s$ = 1 µm and the yarn separation distance is $D_p$ = 5 µm. The spectrally integrated transmittance ($\tau_c$) and reflectance ($\rho_c$) is shown in each plot weighted by the Planck's distribution assuming a body temperature of 33.9°C (93°F). For $D_f$ = 10 µm, the material volume per unit depth for a single yarn is 4870 µm² for $D_y$ = 100 µm, 1492 µm² for $D_y$ = 50 µm, and 550 µm² for $D_y$ = 30 µm. For $D_y$ = 30 µm, the material volume is 373 µm² for $D_f$ = 5 µm and 136 µm² for $D_f$ = 1 µm. The optical properties of the ITVOF are calculated for the wavelength range from 5.5 to 24 µm, which will provide a conservative estimate of the total transmittance and the reflectance.

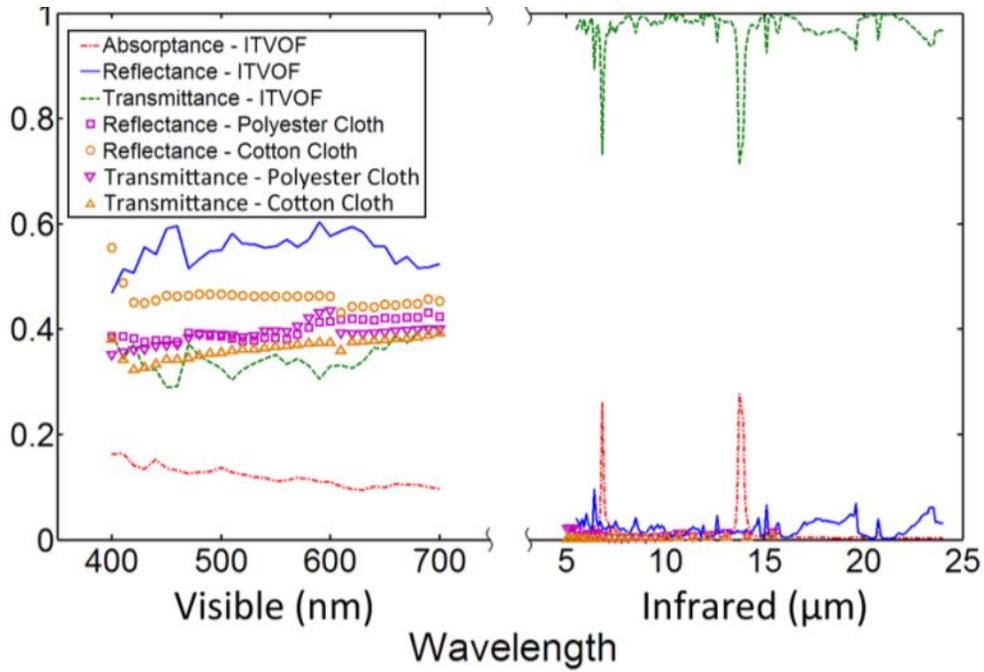

Figure 7: Theoretical results for the visible and IR wavelength range highlighting the contrast in optical properties needed for an ITVOF. These results correspond to the case of $D_f$ = 1 µm, $D_y$ = 30 µm, $D_s$ = 1 µm, and $D_p$ = 5 µm. For comparison, the experimentally measured reflectances and transmittances of cotton and polyester cloths are also shown.

# For Table of Contents Use Only

# Infrared-Transparent Visible-Opaque Fabrics for Wearable Personal Thermal Management

Jonathan K. Tong[1,‡], Xiaopeng Huang[1,‡], Svetlana V. Boriskina[1,*], James Loomis[1], Yanfei Xu[1], and Gang Chen[1,*]

[1]Department of Mechanical Engineering, Massachusetts Institute of Technology, Cambridge, MA 02139

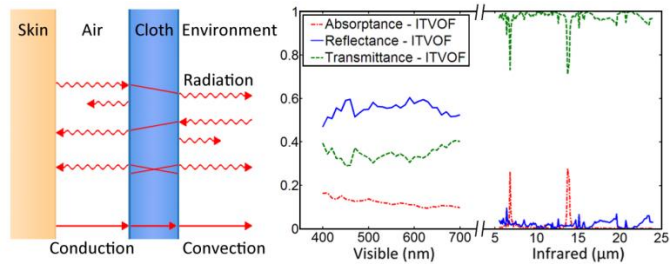

## Synopsis

Personal cooling technologies locally control the temperature of an individual rather than a large space, thus providing personal thermal comfort while supplementing cooling loads in thermally regulated environments. This can lead to significant energy and cost savings. In this study, a new approach to personal cooling was developed using an infrared-transparent visible-opaque fabric (ITVOF), which provides passive cooling via the transmission of thermal radiation emitted by the human body directly to the environment. Here, we present a conceptual framework to thermally and optically design an ITVOF. Using a heat transfer model, the fabric was found to require a minimum infrared (IR) transmittance of 0.644 and a maximum IR reflectance of 0.2 to ensure thermal comfort at ambient temperatures as high as 26.1°C (79°F). To meet these requirements, an ITVOF design was developed using synthetic polymer fibers with an intrinsically low IR absorptance. These fibers were then structured to minimize IR reflection via weak Rayleigh scattering while maintaining visible opaqueness via strong Mie scattering. For a fabric composed of parallel-aligned polyethylene fibers, numerical finite element simulations predict 1 μm diameter fibers bundled into 30 μm yarns can achieve a total hemispherical IR transmittance of 0.972, which is nearly perfectly transparent to mid- and far-IR radiation. The visible wavelength properties of the ITVOF are comparable to conventional textiles ensuring opaqueness to the human eye. By providing personal cooling in a form amenable to everyday use, ITVOF-based clothing offers a simple, low-cost solution to reduce energy consumption in HVAC systems.